\documentclass[prb,twocolumn,showpacs,amsmath]{revtex4}
\usepackage{dcolumn}
\usepackage{bm}
\usepackage{graphicx}

\begin{document}
\title{Crossover from injection to tunneling conduction mode and associated magneto-resistance in a single $Fe_{3}O_{4}$(111)/$Alq_{3}$/Co spin-valve device}
\author{P. Dey}
\author{R. Rawat}
\author{S. R. Potdar}
\author{R. J. Choudhary}
\author{A. Banerjee}
\affiliation{UGC-DAE Consortium for Scientific Research (CSR), University Campus, Khandwa Road, Indore 452 001, India}

\begin{abstract}
We demonstrate interface energy level engineering, exploiting the modification in energy band structure across Verwey phase transition of $Fe_{3}O_{4}$ electrode, in a $Fe_{3}O_{4}$(111)/$Alq_{3}$/Co vertical spin-valve (SV) device. Experimental results on device characteristics (\textit{I-V}) study exhibit a transition in conduction mode from carrier injection to tunneling across Verwey transition temperature ($T_{V}$) of $Fe_{3}O_{4}$ electrode. Both giant magneto-resistance (GMR) and tunneling MR (TMR) have been observed in a single SV device as a function of temperature, below and above $T_{V}$, respectively. Appearance of GMR, accompanied by injection limited natural Schottky-like \textit{I-V} characteristics, provide evidences of spin injection at electrode/$Alq_{3}$ interface and transport through molecular orbitals in this SV device. Features of TMR exhibit significant differences from that of GMR. This is due to the dominant hyperfine-field interaction in the multi-step tunneling regime. We have achieved room-temperature SV operation in our device. A phenomenological model for device operation has been proposed to explain the transition in the conduction mode and associated MR features across $T_{V}$. We propose that the tuning of charge gap at Fermi level across Verwey transition due to charge ordering on the octahedral iron sites of $Fe_{3}O_{4}$ results in a corresponding tuning of conduction mode causing this unique cross over from GMR to TMR in this ferrite-based organic SV.
\end{abstract}
\pacs{85.75.Mm, 81.07.Pr, 72.25.Hg, 75.47.De}

\maketitle
\section {Introduction}

Spin injection, transport and dynamics in interesting surface and interface system, abbreviated as ``spintronics" \cite{Fert} is a fascinating branch of electronics in which the spin degree of freedom of the electrons, along with its charge, is manipulated and controlled to yield a desired outcome where a close relation is witnessed between fundamental scientific activities and technological breakthroughs. A common approach to study spin injection and transport is the observation of magneto-resistance (MR) in spintronics device. The most fundamental form of a spintronics device is spin valve (SV) structure, which, in its preliminary form, consists of two ferromagnetic (FM) electrodes, having well separated magnetic switching fields and a non magnetic spacer layer. Significant difference in resistance could be observed between the parallel and anti-parallel magnetization configuration of FM electrodes of the device attained by applied magnetic field. In case the metallic spacer layers are used, the spin polarized current is injected into and transported through the spacer layer, the observed effect is generally referred as giant MR (GMR) \cite{Fert}. On the other hand, in case of thin insulting spacer layer, spin polarized current tunnel through the insulating layer, hence the observed effect is referred to as tunneling MR (TMR) \cite{Moodera}. Although, proposal for incorporating semiconductor spacer layer is attractive owing to the possible implementation in spintronic logic devices, spin injection becomes challenging because of the conductivity mismatch problem \cite{Schmidt}. Full potential of spintronics could be realized in devices exploiting spin polarized carrier injection in the semiconductor itself. However, coherent spin transport over nanometer scale has not been achieved in both normal metal and semiconductors \cite{Zutic}.

In the early 2000s', an entirely new concept came into the scenario to incorporate organic films into spintronics devices. In general, $\pi$-conjugated organic semiconductors (OSCs) have attracted great deal of attention to researchers and several organic electronic devices have already been demonstrated such as organic light emitting diodes (OLEDs) \cite{Friend}, single-molecule spin switches \cite{Liu}, organic field effect transistors (OFET) \cite{Xiaoran} etc. With respect to the study concerning spin injection in semiconductor structures, it has been found that incorporation of OSC in spintronics will be particularly advantageous for the injection of spin-polarized currents \cite{Naber, Sanvito}. The prospective of OSCs include the ease of fabrication, versatile structural modification, low weight, flexibility, chemically interactive bottom-up fabricated electronics, low production costs, a long spin relaxation times in carbon-based compounds and finally their ability to be integrated in hybrid organic-inorganic devices \cite{Friend, Krinichnyi, Rocha}. Among them, robust spin coherence in most OSCs, which signifies spin propagation over longer distances than in conventional metals and semiconductors, makes them appealing for implementation in spin injection applications and efficient devices. Spin relaxation time is high, in the $10^{-5}$ to $10^{-7}$s range \cite{Harris, Lebedev, Dediu} and could be enhanced by proper chemical engineering by several orders of magnitudes; generate such an intense interest in OSCs \cite{Urdampilleta}. Effective spin diffusion length in OSC is estimated to be $\sim$ 10 nm from muon spin rotation \cite{Drew} and photoemission \cite{Cinchetti} techniques. Some groups have even demonstrated propagation of spin polarization to distances exceeding 100 nm even at room temperature \cite{Barbanera, Xiong, Majumdar, Hueso}. This appreciably weak spin decoherence, i.e., long spin life-times is attributed to intrinsically low spin-orbit coupling due to the low mass of atoms involved and weak hyperfine interactions in organic materials. Furthermore, flexible structure due to weak intermolecular bonds and the numerous chemical synthesis routes of new molecules offer a chemically tailored organic based spintronics devices. Integration of molecular electronics into spintronics, usually referred as ``molecular spintronics", suggests the ordered molecular films as ideal candidates in replacing insulating/semiconducting layers in hybrid spintronic devices \cite{Wolf, Zutic}.

In this direction, pioneering SV effect up to 30 $\%$ has been reported by Dediu {\it et al.} \cite{Barbanera} in spin injection studies on sexiphenyl, making use of $La_{0.7}Sr_{0.3}MnO_{3}$ (LSMO) electrodes in lateral SV device. However, first clear indication of a reasonable and negative SV effect in hybrid vertical organo-metallic devices has been demonstrated by Xiong {\it et al.} \cite{Xiong} in case of spin transport in tris[8-hydroxyquinoline]aluminium ($C_{27}H_{18}N_{3}O_{3}Al$) ($Alq_{3}$) organic molecule with  LSMO and Co, used as bottom and top electrodes, respectively. Although, SV effect was measured in thin layers of OSCs ($<$ 10 nm) as TMR \cite{Santos} without any confusion; the reports \cite{Barbanera, Xiong} on GMR with thick $Alq_{3}$ OSC layers ($>$ 100 nm) have been very controversial \cite{Shim, Vinzelberg, Jiang}. These studies have invoked considerable attention in $Alq_{3}$ based SV devices \cite{Xiong, Hueso, Santos, Vinzelberg, Osterbacka, Xu, Dediu1, Wang}. However, efficient spin-polarized carrier injection and transport in OSC films are rather complicated and the mechanism of such spin-dependent phenomena through molecular media is not completely understood. For instance, both spin injection and spin tunneling transport is possible in this kind of organic based SV devices and a clear distinction between these two mechanisms is required for the proper analysis of SV effect. Study in this direction has been carried out as a function of organic layer thicknesses \cite{Yoo, Lin}. In case of tunneling devices, they are characterized by zero residence time of charge/spin in the organic barrier and exhibits large MR values and thereby offer the functionality as 0-1 switching elements or as magnetic sensors \cite{Dediu}. On the other hand, injection devices are characterized by a finite lifetime of the generated spin polarization in OSC and have potential for the possible implementation as spin manipulation, control of the exciton statistics in the OSC etc \cite{Dediu}.

Among all reported $Alq_{3}$-based SV devices, half metallic manganites (LSMO) electrodes with nominally 100 \% spin polarization demonstrated the most efficient spin injection exhibiting pronounced GMR effect \cite{Barbanera, Xiong, Vinzelberg, Dediu, Yoo}. However, LSMO operates only at quite low temperature thus operational temperature limitation becomes one of the major factors in $Alq_{3}$- based spintronic devices employing LSMO and Co electrodes, without any interface engineering. It is well known that alignment of the energetic position of the highest occupied molecular orbitals (HOMO) or the lowest unoccupied molecular orbitals (LUMO) levels of the OSC, relative to the Fermi level ($E_{F}$) of the FM electrodes is crucial for efficient carrier injection and extraction \cite{Dediu}. Interface engineering or tuning of band alignment of the electrodes with respect to molecular orbitals has been a continuous approach in order to modulate or optimize the SV device properties. However, up to this point interface engineering has been done either by improving the deposition condition or by introducing additional polar/seed layer in between electrode and molecular layer \cite{Santos, Dediu1, Schulz, Jang}. Here, our approach is a bit `intrinsic', for we intend to incorporate a material as an injector electrode that itself would exhibit energy band tuning triggered by any external thermodynamic parameter. Our judicious choice with this objective was $Fe_{3}O_{4}$ (111). It is well known that stoichiometric bulk magnetite undergoes a phase transition across Verwey transition temperature ($T_{V}$) $\approx$ 122 K \cite{Verwey, Verwey1}, which is associated with a drastic change in electrical conductivity, magnetic property and structure of $Fe_{3}O_{4}$. Among various concepts, charge ordering on the octahedral iron sites of $Fe_{3}O_{4}$ causing modification in energy band structure and thereby inducing Verwey transition, is widely supported by various groups \cite{Park, Tokura}. Here, we present our study on $Alq_{3}$- based vertical SV structure employing $Fe_{3}O_{4}$ (111) and Co electrodes, i.e., we are introducing ferrite electrode instead of manganites, as conventionally used in oxide based organic SVs. We are interested to see the effect of Verwey phase transition of $Fe_{3}O_{4}$ on the SV device properties. $Fe_{3}O_{4}$, having $T_{C}$ = 858 K, is predicted to be a half-metal, i.e., having large spin-polarization at the Fermi level at room temperature. Moreover, $Fe_{3}O_{4}$ exhibits an electrical resistivity of the same order of magnitude as semiconductor hence the conductivity mismatch problem would be less as compared to LSMO or other spin injectors. A very recent study \cite{Pratt} exhibits the energy-level alignment at $Fe_{3}O_{4}$/$Alq_{3}$ interfaces, which confirms that $Fe_{3}O_{4}$ could be incorporated as an efficient spin-injector into OSC in organic based spintronic devices. In fact, few studies have been devoted in this direction \cite{Pratt}, however, no conclusive experimental evidence of SV effect has been obtained so far. In this paper, we present experimental results on $Fe_{3}O_{4}$(111)/$Alq_{3}$/Co SV device, exhibiting a crossover in conductivity mode from carrier injection to tunneling across Verwey transition temperature of $Fe_{3}O_{4}$ electrode. Both GMR and TMR have been obtained in a single SV device as a function of thermodynamic parameter temperature. Thus, a drastic tuning in SV device properties can indeed be achieved triggered by the Verwey phase transition of $Fe_{3}$$O_{4}$ electrode. Effect of hyperfine-field interaction, which is weak in injection regime, is found to be dominant in tunneling regime, resulting in modification of TMR features. TMR SV effect is found to persist up to room temperature in our device. A phenomenological model for device operation is proposed to explain the observed transition in the conduction mode and the associated MR, across Verwey transition temperature of $Fe_{3}O_{4}$ electrode.

\maketitle\section{EXPERIMENTAL DETAILS}

We have fabricated $Alq_{3}$ based vertical SV device in an \textit{ultra}-high vacuum deposition system \cite{ebeam}, employing electron-beam evaporation technique with a base pressure in the range of $10^{-9}$ mbar. A sublimed grade 99.995\% trace metal basis $Alq_{3}$ (Sigma-Aldrich) was used. In order to realize the SV structure for transport measurement, we have used a mechanical shadow mask, which is qualified for changing under \textit{ultra-high} vacuum conditions in a continuous vacuum process. Deposition set up consists of multicrucible electron evaporator. General structure of the SV is as follows: bottom FM electrode ($Fe_{3}O_{4}$)/organic spacer ($Alq_{3}$)/top FM electrode (Co)/cap layer (Au). We have used the substrate having size of 10 x 10 $mm^{2}$. The bottom electrode, i.e., $Fe_{3}O_{4}$ (111) thin film ($\sim$800 \AA) was deposited on Si (100) substrate from $Fe_{3}O_{4}$ target by pulsed laser deposition technique using KrF excimer laser source ($\lambda$ = 248 nm). Details of film deposition and characterizations were published elsewhere \cite{PLD}. After methanol cleaning $Fe_{3}$$O_{4}$ films were loaded in the e - beam deposition set-up where the SV structure was completed by the deposition of $Alq_{3}$ and the counter electrodes Co and Au.

In the e-beam evaporation set up, one of the crucible positions was modified for the evaporation of the organics by introducing an additional graphite boat. The boat is covered, except a small orifice, ensuring that e-beam is hitting the boat without being in contact with organic materials inside it. In this way, we have successfully evaporated and deposited organic material through indirect thermal evaporation process at room temperature. It is well known that room temperature deposition of organic molecules on oxides surfaces provides smooth and uniform growth and morphologically stable amorphous organic film with molecularly flat surfaces \cite{Dediu1}. We could evaporate in our system with the same electron gun $Alq_{3}$ layer, Co and Au overlayer. The deposition rate of $Alq_{3}$ was kept at 1.0 - 3.4 \AA/sec and the deposition rate of Co and Au were 0.1 - 0.2 \AA/sec. Larger source-substrate distance for the Co deposition was supposed to reduce the damage and inter diffusion in the organic layer during top electrode deposition. Film thickness was monitored during deposition by a calibrated quartz crystal monitor. The thickness of Co and Au layer was kept both 100 \AA\ and for $Alq_{3}$ layer was 1100 \AA. The device was rectangular bar shaped with junction area $\approx$ 2 x 2 $mm^{2}$.

Electrical and MR measurements were carried out using standard four probe method in a home made resistivity/MR set up alongwith 8 Tesla Oxford Superconducting Magnet system. This system provides measurements in the temperature range from 4.2 - 300 K. In the present case, typical device resistance below 100 K exceeds instrumental limit and hence we have restricted our measurements down to the lower limit of temperature of $\sim$ 100 K. MR measurements were carried out in constant current mode with external magnetic field along the long axis of the bar shaped electrode and parallel to the layers. Magnetic hysteresis loops were measured with a Quantum Design Superconducting quantum interference device (SQUID) magnetometer.

\maketitle\section{EXPERIMENTAL RESULTS AND DISCUSSIONS}

\maketitle\subsection{Current-Voltage (I-V) Characteristics}

While the spin diffusion length in OSC is estimated to be $\lambda$$_{S}$ $\sim$ 10 nm \cite{Drew, Cinchetti}; GMR observed in SV devices employing OSC thickness layer $\sim$ 100 nm \cite{Barbanera, Xiong, Vinzelberg, Dediu1} invokes much controversy. A complete understanding and assured signatures of spin injection and transport in OSC layer is still lacking due to lack of comprehensive device characteristic study \cite{Szulczewski}. For this reason, we have performed \textit{I-V} study to shed light on how carriers are mediated through the $Alq_{3}$ organic layer depending on bias and temperature. Figures 1(a) and (b) show \textit{I-V} characteristics, measured on $Fe_{3}O_{4}$(111)/$Alq_{3}$/Co SV device at different temperatures ranging from room temperature (300 K) down to 100 K. In the following discussion, positive voltage means that $Fe_{3}O_{4}$ is the anode and Co is the cathode and negative voltage implies the opposite picture. All the curves are non-linear, asymmetric and exhibit insulating-like behavior with decreasing temperature thus rules out the contribution of possible shorts circuit channels in these devices \cite{Jonsson}. The non-linearity in \textit{I-V} curves implies on first hand approximation the existence of strong interface barriers. Asymmetry in the \textit{I-V} characteristics comes directly from the asymmetry in the device structure and different work functions of the electrodes employed. It is evident from Figs. 1(a) and (b) that there exist entirely two different modes of conductivity for the same SV structure as a function of temperature. The conductivity mode at T = 100 K is completely different than what is observed at temperatures above this. At T = 100 K, \textit{I-V} curves share their features with those typically measured in OLEDs \cite{Lin}. They are characterized by an onset voltage, i.e., barrier potential, below which the current is very small and above which current increases sharply in a nonlinear fashion. Appearance of such onset voltage still could be identified at T = 150 K [inset in Fig. 1(a)] but above these temperatures the characteristic feature of \textit{I-V} curves change completely. Generally, the current conduction mode observed at T = 100 K is described by `activated' carrier injection into the molecular orbitals of OSC at the electrode-OSC interface, followed by `activated' carrier hopping through the molecular orbitals towards opposite electrode, where the extraction of carriers from OSC takes place. On the question of optimal interface state for reaching an efficient spin injection and detection, i.e., good MR, this kind of natural Schottky-like barriers at the interface seems adequate for realization of organic SVs \cite{Dediu, Smith}.

It is evident that the conductivity is much more at positive bias voltage than that of negative voltage thus yielding a very asymmetric \textit{I-V} curve at T = 100 K [Fig. 1(a)]. With the knowledge of the work functions of Co and $Fe_{3}O_{4}$, which are 5 and 5.2 eV, respectively, and the ionization potential of $Alq_{3}$, which is 5.8 eV; we have constructed an energy level diagram [Fig. 2 (a)], which compares the work functions of the electrodes with that of HOMO and LUMO levels of $Alq_{3}$. Schematic illustration of device structure and the measurement geometry is shown in Fig. 2 (b). Energy levels of $Alq_{3}$ are such considered that are relevant for transport, characterized by an energy gap between HOMO and LUMO levels of about 4.6 eV, as shown by a combination of photoemission and inverse photo emission spectroscopy \cite{Jiang, Knupfer}. It should be clarified in this context that the optical gap is 2.7 eV and the rest is contributed by the exciton binding energy \cite{Knupfer}. In many previous literatures \cite{Xiong, Dediu1}, only the optical gap is considered while explaining transport property. However, this is a quite generic feature that the exciton (Frenkel excitons) binding energy in OSCs is larger than in inorganic semiconductors, thus resulting in a larger separation between transport gap and optical gap than in inorganic semiconductors \cite{Knupfer}. It is well known that the presence of interfacial dipoles at the electrode/$Alq_{3}$ interface often give rise to vacuum level offset, which generally shifts the non-interacting equilibrium energy levels of molecules towards lower energy range. The reported interfacial barriers of $Alq_{3}$ with Co \cite{Zhan} and $Fe_{3}O_{4}$ \cite{Pratt} are 1.4 and 1.2 eV, respectively. With all these considerations, it follows that for Co and $Fe_{3}O_{4}$, the hole injection barriers to HOMO level of $Alq_{3}$ are 2.2 and 1.8 eV and the electron injection barriers to that of LUMO level are 2.4 and 2.8 eV, respectively. At T = 100 K, during the positive polarity of bias voltage current rises rapidly when V $>$ 2 eV [Fig. 1 (a)]. This is the case when $Fe_{3}O_{4}$ is the anode, where holes may be injected from $Fe_{3}O_{4}$ and electrons may be injected from Co. Since the energy barrier for electron injection from Co ($\sim$ 2.4 eV) is higher than the barrier for hole injection from $Fe_{3}O_{4}$ ($\sim$ 1.8 eV), the hole current is expected to dominate. In fact, $Fe_{3}O_{4}$ was already shown to enhance hole injection in OLEDs acting as an anodic buffer layer \cite{Zhang}. In a similar way, hole current is expected to dominate when Co is the anode as well. Thus current in the device is predominantly carried by holes traversing the HOMO. However, very low conductivity associated with negative polarity of bias voltage [Fig. 1 (a)], i.e., when Co is the anode, is justified by considering reasonably large both hole ($\sim$ 2.2 eV) and electron ($\sim$ 2.8 eV) injection barrier from Co and $Fe_{3}O_{4}$ electrodes, respectively.

We understand that carrier injection at the electrode-$Alq_{3}$ interface takes place via thermionic and/or bias field emission as the carriers must overcome a significant energy offset between electrode work function and HOMO levels of $Alq_{3}$. In thermionic field emission mechanism, conduction carriers tunnel through the potential barrier at the interface. We have shown in Fig. 3(a) the \textit{log I vs. log V} plots at T = 100 and 150 K, where with increase in bias field curves for two different temperatures asymptote to the same slope. This is a typical characteristic for charge injection via field emission at electrode-organic interface \cite{Yoo}. Unlike injection mode in OLED where at very low bias conductivity is negligible \cite{Lin}, in this present case carrier injection into $Alq_{3}$ is found to be non trivial even at very low bias. This is possible only as a temperature activated process, thus, as expected, this current conduction mode is strongly temperature dependent.

Interestingly, at higher temperatures (T $>$ 150 K) [Fig. 1 (b)] \textit{I-V} curves are clearly characteristics of tunneling case resembling those seen in diversified inorganic \cite{Moodera, Fan} and organic magnetic tunnel junctions (MTJs) \cite{Santos}. It is well known that the low-bias \textit{I-V} curves of a tunnel junction obey the relationship, [I = (V + 0.625 $V^{3}$)/1330] \cite{Simmons}. To verify this in our case, \textit{I-V} curves are numerically differentiated resulting in differential conductance curves. Figure 3(b) exhibits differential conductivity (\textit{dI/dV}) versus bias voltage at T = 100 and 250 K. At T = 250 K, differential conductance curve is found to be near-parabolic typical of tunneling conduction \cite{Moodera, Santos, Fan}. However, at T = 100 K the characteristics of conductance curve is totally different from that observed at T = 250 K and clearly exhibits the beginning of the injection mode at a bias of $\approx$ 2 V. These differential conductivity (\textit{dI/dV}) curves at both T = 100 and 250 K exhibit a sharp dip at zero bias, which can be clearly seen from the inset in Fig. 3(b) [at T = 250 K]. Presence of such a dip at conductance curve is known as zero bias anomaly. Diffusion of magnetic impurities into the barrier was suggested to be one of the prime reasons for this anomaly \cite{Appelbaum}. Hence the observed sharp dip in conductance curve in this case implies that the barrier and interfaces in our SV have top electrode Co inclusion, which was also widely found in previous literatures in case of such Co top electrode based SV structure \cite{Xiong, Vinzelberg}. Moreover, as one of the generic feature of tunneling conduction \textit{I-V} curves are characterized by weak temperature dependence \cite{Lin} [Fig. 1(b)] compared to low temperature injection mode, which is strongly temperature dependent [Fig. 1(a)]. In this context, we should mention the conduction process, achieved through phonon assisted field emission \cite{Yoo}. In this conduction process, although carriers are injected into the organic layer, since carrier injection can take place even at very low bias, \textit{I-V} curves in some case could resemble to that of tunneling mode as identified in an earlier work \cite{Yoo}. However, in that case conduction was found to be highly temperature dependent and hence could not account for our observation of significantly weak temperature dependence of conduction mode at higher temperature regime [Fig. 1(b)]. Thus we are justified to assign this high temperature conduction mode as achieved through carrier tunneling process. It comes out that there is a drastic change in conductivity mode in a single SV device from injection to tunnel mode, as a function of thermodynamic parameter `temperature'. We will explain this in our discussion part.

\maketitle\subsection {Magnetoresistance (MR)}

In this section, we present spin-dependent transport measurements carried out on $Fe_{3}O_{4}$(111)/$Alq_{3}$/Co SV structure of constant lateral size of 2 x 2 $mm^{2}$. The electrical properties are reproducible over days. Figure 3 (c) shows the resistance variation (at a constant current) as a function of magnetic field, where magnetic field is applied in-plane of the sample and parallel to current, i.e., longitudinal MR. Clear MR hysteresis loops have been observed at all temperatures from 100 to 300 K. Remarkably, we obtained room temperature MR [Fig. 3 (c)] in this large area, thick $Alq_{3}$ barrier SV device. Given the fact that the difference in the electrode workfunctions gives rise to a built-in potential ($V_{bi}$) of $\approx$0.3 V across $Alq_{3}$ layer at zero applied bias ($V_{appl}$), effective voltage drop ($V_{eff}$) across the organic barrier should be $V_{eff}$ $\approx$ $V_{appl}$ - $V_{bi}$. Current is set such that $V_{appl}$ remains around 0.3 V, thereby confining the effective bias voltage for MR measurements in a very low range. This is motivated from the fact as identified from previous studies that the MR effect is very strong at low bias voltage and reduces with increasing voltage \cite{Xiong}. Concentrating on the MR behavior in the carrier injection regime, i.e., at T = 100 K, we found clear and reproducible positive MR, as can be clearly seen in upper panel of Fig. 4(a). This is in contrast with the results obtained in the case of similar millimeter square large SVs and thick $Alq_{3}$ layers in the 130/250 nm range, where negative MR has been routinely obtained since the first result \cite{Xiong, Hueso, Vinzelberg, Osterbacka, Xu, Dediu1, Wang}. It should be noted that some fine structure is visible in the antiparallel state, similar to that observed in EuS-based filters \cite{LeClair}. This feature can be associated with the presence of domains and/or domain walls in the junction area. Observation of this SV effect, accompanied by natural Schottky-like \textit{I-V} characteristics [Fig. 1(a)] indicates efficient spin polarized carrier injection, transport and extraction, which are the main ingredient of spintronics, could be successfully achieved in this ferrite-based SV device. Further, we have also ruled out tunneling anisotropic MR as origin of this observed SV effect by performing transverse (current perpendicular to magnetic field) MR measurements (not shown). It should be noted that MR remains positive irrespective of applied voltage.

It is well known that in order to observe SV effect both top and bottom FM electrodes should have well-separated magnetic coercivities. For the applied external magnetic field (\textit{H}) in between the coercivities of two FM electrodes, the magnetization orientations between them are anti-parallel to each other. On the other hand, when \textit{H} is greater than both of the coercivity values at a given temperature, magnetizations are parallel to each other. At T = 100 K, steps of the MR curve well correspond to that of in-plane magnetic coercivities of the two FM electrodes at the same temperature as is evident from \textit{dM/dH} plot in middle panel of Fig. 4(a). This can also be clearly seen from the smoothened resistance versus applied \textit{H} curve, as shown in inset of lower panel of Fig. 4(a). Here higher and lower switching fields correspond to that of $Fe_{3}O_{4}$ and Co electrodes, respectively. Thus the observed positive hysteresis loop in MR at T = 100 K straight forwardly implies low-resistance state corresponding to parallel and high-resistance state to that of antiparallel magnetization configuration of two FM electrodes. Interestingly, similar distinct steps in MR curves were not found at higher temperatures. At T = 150 K, such steps in MR curves still can be identified [Fig. 3(c)]. However, for T $>$ 150 K features associated with MR curves change completely [Fig. 3(c)], where the conduction mode changes from carrier injection to tunneling, as identified from \textit{I-V} curves (discussed in section A). This is evident from upper panel of Fig. 4(b), where at T = 250 K MR peaks at \textit{H} $\approx$ 250 Oe, which corresponds to the magnetization switching of $Fe_{3}O_{4}$ electrode at that temperature, followed by a sharp decrease of MR with increase in field. However, at low field regime MR shows a rather continuous increase with rise in field from zero value. More importantly, for both direction of field resistance starts to increase even for parallel alignment of magnetizations (at low field regime). Noteworthy feature is that magnetic coercivities between $Fe_{3}O_{4}$ and Co electrodes are still well separated ($\sim$ 200 Oe) at this temperature [middle and lower panel of Fig. 4(b)] with Co electrode switches distinctly at very low field. Thus it comes out that at high temperature tunneling regime, distinct MR switching corresponding to the magnetization switching of Co electrode could not be obtained. We will address this behavior in the next section.

There is still lacking of well-defined models for spin injection and transport in OSCs. However, MR measurements could provide qualitative information. By analyzing the amplitude of MR, its temperature dependence and precise analysis of its sign could possibly explore several fundamental physical aspects governed by the interfacial properties of OSCs and metallic/oxide electrodes. In our discussion, first we consider the sign of MR. As recognized in previous studies in case of thick (130 - 200 nm range) $Alq_{3}$ layers, the real thickness should actually be reduced by up to 100 nm because of top electrode diffusion in organics, thus leaving the effective junction thickness only few tens of nanometers \cite{Xiong}. In this framework, the sign of MR was generally analyzed by Jullière model \cite{Julliere}, which in the case of negative MR assign opposite spin polarization to the two electrodes. This, in turn implies that the polarity of injected spins from injecting electrode to OSC and extracted spins in extracting electrode from OSC, is reversed \cite{Dediu1}. It was also said that for the device where spin extraction occurs below the Fermi level, minority carrier extraction may be more efficient, thus yielding negative MR \cite{Schulz}. On the other hand, it was also shown that for devices where both spin injection and extraction occurs near the Fermi level, injected electrons with majority spin from one electrode are preferred to be extracted in another, thus producing positive MR \cite{Jang}. However, it has been shown that both positive and negative MR  could be obtained in a single device as a function of applied voltage \cite{Vinzelberg}. Furthermore, for the same interface system LSMO/$Alq_{3}$, both positive and negative MR has been reported depending on whether they are locally probed in thin tunnel barrier \cite{Barraud} or has been observed for thicker and larger ones \cite{Dediu1}. Following the spin transport model proposed by Barraud {\it et al.} \cite{Barraud}, the specific spin dependent hybridization of the molecular orbitals at the electrode-molecule interface induced polarized states in the first molecular layer at the electrode surface and thereby control the effective polarization at the interface. This, in turn accordingly modulates the sign of injected/extracted spins and hence that of MR. It also follows from their model that for weak coupling between electrode and OSC, this effective polarization is positive and even could be levered. Interestingly, for strong coupling between electrode and OSC, the effective interfacial polarization could possibly be inverted. More subtle point is that this inversion should not happen for all of the interfacial molecular levels. Instead this effect is restricted to only for those orbitals, which are close enough to electrodes Fermi level and governing injection step. It followed from this model that polarization at LSMO/$Alq_{3}$ is inverted thus implying strong coupling at this interface, whereas for Co/$Alq_{3}$ it is positive, which has also experimental supports from literature \cite{Santos, Dediu1}. Strong polaronic character of LSMO could play significant role in the strong coupling between LSMO and OSC \cite{Dediu}.

From this discussion, we would try to explore some possible attributes about the hybridization at electrodes/$Alq_{3}$ interface and the spin/charge transport in our $Fe_{3}O_{4}$(111)/$Alq_{3}$/Co SV device. From the observed positive sign of MR in this SV device [Fig. 3 (c)], we can conclusively state that both electrodes must inject and extract same spins, i.e., spin-majority holes at electrode/$Alq_{3}$ interface in the carrier injection regime. Thus, similar to Co/$Alq_{3}$ interface that is well known to inject spin majority carriers; in this case $Fe_{3}O_{4}$/$Alq_{3}$ interface is also revealed to inject/extract majority spin carriers. For this to happen there are two possibilities either, i) unlike LSMO, the effective interfacial polarization of $Fe_{3}O_{4}$/$Alq_{3}$ interface is positive; or ii) carrier injection/extraction at $Fe_{3}O_{4}$/$Alq_{3}$ interface takes place through localized states, which are off-resonance with the Fermi level of $Fe_{3}O_{4}$. Given the fact that similar to LSMO, $Fe_{3}O_{4}$ is also having strong polaronic character; strong coupling at $Fe_{3}O_{4}$/$Alq_{3}$ interface is expected that might invert interfacial polarization close to $Fe_{3}O_{4}$ Fermi level. In this scenario, the concept of off-resonance carrier injection/extraction through localized states at $Fe_{3}O_{4}$/$Alq_{3}$ interface is highly apt.

One subtle feature observed in our SV structure is that the SV effect is less pronounced during the first magnetic field sweeping between positive to negative saturation magnetic field (where both electrodes attain magnetic saturation), as is evident in Fig. 5. However, with recycling the magnetic field the feature becomes more prominent. After two/three successive magnetic field sweeping between positive to negative saturation magnetic field, clear characteristics related to SV effect becomes established and no more changes with further recycling the magnetic field is identified. Interestingly, no such training effect is observed in case of magnetization hysteresis loop associated with this SV structure. It is well known that it is the surface magnetic property rather than bulk magnetism that play decisive role for an electrode to act as an efficient injector \cite{Dediu1}. As recognized above that the two electrodes inject/extract spin-majority holes, there is possibly a spin-majority electron accumulation at the cathode interface, which was also experimentally evidenced in a different organic based SV \cite{Schulz}. This may in turn cause an enhancement of interface magnetization of the electrodes. Thus a cumulative effect is expected at the initial stage where the circulation of spin majority hole current through the device causes enhancement of interface magnetization, which consequently results in more pronounced SV effect.

\maketitle\subsection {Discussions}

In this section, we will address the change in conduction mode from carrier injection to tunneling as indicated from \textit{I-V} measurements (Section A) and the associated change in MR features (Section B) in a single SV device, as a function of temperature. As already mentioned, stoichiometric bulk magnetite under goes a phase transition between a high temperature ``bad metal'' phase and a low temperature insulating phase across $T_{V}$ $\approx$ 122 K \cite{Verwey, Verwey1}. In our case, we found from both magnetic and electronic-transport measurements $T_{V}$ to be around 120 K (not shown), thereby confirming good stoichiometry of this $Fe_{3}O_{4}$ electrode retaining bulk property. The concerned temperature range of 100 K $\leq$ T $\leq$ 150 K, across which the observed transformation of device transport takes place, signifies the Verwey transition of $Fe_{3}O_{4}$ electrode as origin of this. Widely accepted explanation behind this Verwey phase transition is the charge ordering on the octahedral iron sites of $Fe_{3}O_{4}$ \cite{Park, Tokura}. This charge ordering opens an energy gap of about $\sim$ 0.15 eV at Fermi level as found from photoemission studies \cite{Park}. Activated carrier transport in $Fe_{3}O_{4}$ could also be justified in this picture. Above $T_{V}$, with the closure or even with decrease of this energy gap \cite{Tokura}, all the charge carriers take part in the conduction process. It has also been shown that the jump in conductivity across $T_{V}$ is due to the jump in the number of charge carriers and that there is no sudden change of mobility of charge carriers \cite{Kuipers}. More subtle point is that both above and below $T_{V}$, the mobility and hence the conduction mechanism of charge carriers is similar and that is through phonon-assisted hopping \cite{Kuipers, Fursina}. Small polaron formation could be at the origin of this thermally activated behavior of conduction electrons.
These transport characteristics of $Fe_{3}O_{4}$, associated with Verwey transition and the experimental results, obtained in this ferrite based SV device, allow us to propose a simple phenomenological model [Fig. 6], which can describe the device transport properties. At first hand, there are two possible path ways for the device current, one is tunneling through the optical gap and another is carrier injection and transport through the HOMO levels of $Alq_{3}$. In this case, the tunneling is not the conventional direct elastic tunneling through the OSC. For elastic tunneling to occur through the barrier, the thickness of the OSC should be restricted to only $\sim$ 2-3 nm. Given that the thickness of $Alq_{3}$ layer in our case $\sim$ 120 nm, we refer the tunneling here as multistep tunneling through the defect states in the optical gap of the $Alq_{3}$ organic barrier \cite{Yoo}. Existence of such defect states, both of extrinsic and/or intrinsic origin, in the $Alq_{3}$ band gap was evidenced in many earlier works including that in state-of-art $Alq_{3}$ based nanojunction \cite{Barraud}. However, these defect states could not essentially constitute a direct charge path way. Either thermal or electric field activation could possibly elevate the carriers from one defect state to others thereby constituting charge path way towards opposite electrode. Thus, it is straight forward to justify that this tunnel mode could suitably take place at higher temperature regime and diminishes at lower temperature.
\par
First, let us consider the case for T$\leq$$T_{V}$ [Fig. 6(a)]. As described above, there is a formation of charge gap at the Fermi level of $Fe_{3}O_{4}$ that shifts the energy level responsible for carrier transport even at higher binding energies with respect to the vacuum level. Thus the energetic offset between that energy level of $Fe_{3}O_{4}$ electrode and HOMO level of $Alq_{3}$ should be reduced. This, in turn makes the condition for carrier injection from $Fe_{3}O_{4}$ to HOMO level of $Alq_{3}$ more favorable. This is indicated by thick continuous line in Fig. 6(a). On the other hand, since both the numbers of charge carriers and temperature are pretty less, the tunneling, mediated through defect states at the optical gap of $Alq_{3}$ barrier, is quite less probable. This is indicated by thin dashed line in Fig. 6(a). Thus at T$\leq$$T_{V}$, the device current is governed by carrier injection and transport through molecular levels of $Alq_{3}$. This is, in fact the case as explored from \textit{I-V} results [Fig. 1(a)] in our ferrite based SV device. As per expectation, the SV effect in this case is exhibiting MR switching in accordance to that of magnetization switching of the relevant electrodes. By definition, this observed SV effect below $T_{V}$ could be assigned as GMR.

Now, let us consider the case for T$\geq$$T_{V}$  [Fig. 6(b)]. As described above, there is decrease or even closure of the energy gap, formed at the Fermi level of $Fe_{3}O_{4}$ below $T_{V}$. As a result, the effective work function of $Fe_{3}O_{4}$ reduces with respect to vacuum level. This, in turn enhances the energetic offset between the Fermi level of $Fe_{3}O_{4}$ electrode and HOMO level of $Alq_{3}$. Thus the condition for carrier injection from $Fe_{3}O_{4}$ to HOMO level of $Alq_{3}$ becomes less probable. This is indicated by thin dashed line in Fig. 6(b). However, in this case large number of charge carriers and high temperature could enhance the probability for tunneling conduction through defect states. This is indicated by thick continuous line in Fig. 6(b). Thus at T$\geq$$T_{V}$, according to our phenomenological model the device current should be governed by tunneling conduction through the optical gap of $Alq_{3}$, thereby reproducing \textit{I-V} results [Fig. 1(b)] in our ferrite based SV device. Correspondingly, the observed MR in this high temperature regime (T$\geq$$T_{V}$) could be justifiably assigned as TMR [upper panel of Fig. 4(b)].

As already described in Section B that in this high temperature tunneling regime, MR switching corresponds to the magnetization switching of $Fe_{3}O_{4}$ electrode, however, there is no evidence of MR switching corresponding to that of Co electrode. At first glance, it seemed to be the anisotropic MR contribution from $Fe_{3}O_{4}$ electrode to this high temperature MR of the device. However, in plane transverse MR measurement with current perpendicular to the magnetic field [inset in upper panel of Fig. 4(b)] reveal same characteristic feature to that of longitudinal one [upper panel of Fig. 4(b)] without showing any signature of inversion of MR. This, in turn rules out the possibility of anisotropic MR of $Fe_{3}O_{4}$ electrode as origin of this high temperature MR behavior. Considering the fact that this high temperature tunneling conduction mode corresponds, in physical picture, a multi-step tunneling of charge carriers through localized molecular sites; this is the scenario where hyperfine field interaction plays a significant role \cite{Schoonus}. It has already been reported that charge carriers spent appreciable time ($>$ 10 ns \cite{Bobbert}) on the localized molecular site between two successive hops. Thus the associated spin with the charge carriers could become particularly sensitive to the randomly oriented local hyperfine fields ($H_{hf}$) from the hydrogen nuclei \cite{Schulten, Wohlgenannt}. Given that the $Alq_{3}$ layer thickness is quite large ($\sim$ 120 nm) in the present case; when \textit{H} $\gg$ $H_{hf}$ spins follow \textit{H} straightforwardly, however, for lower values of \textit{H} hyperfine interaction is supposed to take place. Hyperfine interaction is the precession of spins at the molecular site about the local $H_{hf}$, which is basically quasi static in nature and vary randomly in direction from molecule to molecule \cite{Schulten}. Thus this hyperfine field interaction would lead to decoherence of the carrier spin in this low field range that in turn causes the rise in resistivity even in parallel configuration of magnetization of the FM electrodes \cite{Schoonus}. With the similar argument, there should be a decrease in resistance in the anti-parallel configuration of the electrodes at lower field side \cite{Schoonus}. Thus, this description could qualitatively explain our observed MR features [upper panel of Fig. 4(b)] in the tunneling regime. In this context, we should point out that MR switching, corresponding to that of Co electrode, seems to be smeared out by the hyperfine interaction induced resistance increase/decrease in parallel/anti-parallel (in the low field regime) configuration of magnetization of the FM electrodes. It should be noted that in injection regime also, i.e., at T = 100 K, similar hump-like increase in resistance at very low field regime for parallel alignment of magnetizations could be identified [upper panel of Fig. 4(a)]. However, the observed effect is pretty less compared to that in tunneling regime. In the injection regime T $\leq$ $T_{V}$, after carriers get injected into the HOMO levels of OSC from electrode, charge transport take place through $\pi$-delocalized orbitals, constituted by the C atoms of the $Alq_{3}$ molecules. Thus, in that case there is very less probability for hyperfine field interaction between injected spins and hyperfine field of hydrogen nuclei. In fact, electronic transport through $\pi$ delocalized orbitals in case of any OSC, is generally believed to play an important role in deciding long spin relaxation time because of less hyperfine field interaction \cite{Barraud}. Observation of this modulation of hyperfine field interaction with the change in conduction mode, is although a subtle, nevertheless a significant one. For this in turn, further convinced the concept of carrier injection and transport through molecular orbitals in injection limited regime and carrier transport through molecular site by multistep hopping in tunneling regime.

Furthermore, the role of possible Co inclusions in the $Alq_{3}$ barrier, which was also experimentally evidenced by the presence of zero-bias anomaly in conductance curve [inset in Fig. 3 (b)], in determining the behavior of MR is discussed. The magnetic moments associated with these small Co inclusions are expected to be blocked, decided by their anisotropy energy, at very low temperature range and above these blocking temperatures their magnetizations are randomized. Thus, these small Co inclusions could not act as an active counter electrode to $Fe_{3}O_{4}$ and remain inactive at our operating temperature range. Hence, it is the Co layer that only constitutes the active counter electrode to $Fe_{3}O_{4}$.

\maketitle\section{CONCLUSIONS}

In conclusion, we have attempted interface engineering through band alignment tuning at the electrode/$Alq_{3}$ interface, employing the Verwey phase transition of $Fe_{3}O_{4}$ electrode. We have clearly demonstrated a crossover of conduction mode from carrier injection to tunneling across Verwey transition temperature of $Fe_{3}O_{4}$ electrode in $Fe_{3}O_{4}$(111)/$Alq_{3}$/Co SV device. Both GMR and TMR have been observed in a single SV device as a function of thermodynamic parameter temperature with GMR/TMR occurs below/above $T_{V}$. Appearance of GMR effect, accompanied by injection limited natural Schottky-like \textit{I-V} characteristics, indicates with reasonable confidence that efficient spin polarized carrier injection at electrode/$Alq_{3}$ interface, transport through molecular orbitals and finally extraction of spins at extracting electrode, which are the main ingredient of spintronics, could be successfully achieved in this ferrite-based SV device. On the other hand, in the tunneling regime above $T_{V}$, multi-step tunneling of charge carriers through defect states at the optical gap of $Alq_{3}$, results in modification of TMR features due to significant hyperfine-field interaction. Noticeably, TMR SV effect persists up to room temperature in our device. We have proposed a phenomenological model for device operation. Within this model, the concept of charge gap opening and closing at Fermi level across Verwey transition due to charge ordering on the octahedral iron sites of $Fe_{3}O_{4}$, is envisaged to offer this unique scenario of tuning the conduction mode and hence MR in this ferrite based organic SV. Thus our study confirms that with the incorporation of $Fe_{3}O_{4}$ electrode in a SV device, a drastic modulation in SV device properties triggered by the intrinsic Verwey phase transition of $Fe_{3}O_{4}$ electrode, can certainly be achieved. A corresponding tunability of its functionality could also be anticipated. Fabrication of top oxide electrodes has been proposed to circumvent the short circuit problem, as generally found in organic devices with metallic top electrode, due to the much lower diffusivity of oxide species. Use of $Fe_{3}O_{4}$ films for this role is appealing due to the possibility to grow high quality films even at room temperature. In the present case, although we have studied $Fe_{3}O_{4}$ bottom electrode; our experimental data seems to suggest the promising use of $Fe_{3}O_{4}$ as a top electrode in oxide based organic SV.

\maketitle\section{Acknowledgements}

We thank Ridhi Master for preparing $Fe_{3}O_{4}$ substrate and for the helpful discussion on the details of structural and transport properties of the substrate. Sachin Kumar is acknowledged for help in Magneto-Transport measurements. Kranti Kumar Sharma is acknowledged for the Magnetic measurements. We are thankful to Ajay Gupta for helpful discussions regarding film deposition, characterization and x-ray reflectivity.

\begin{figure*}
	\centering
		\includegraphics[width=12cm]{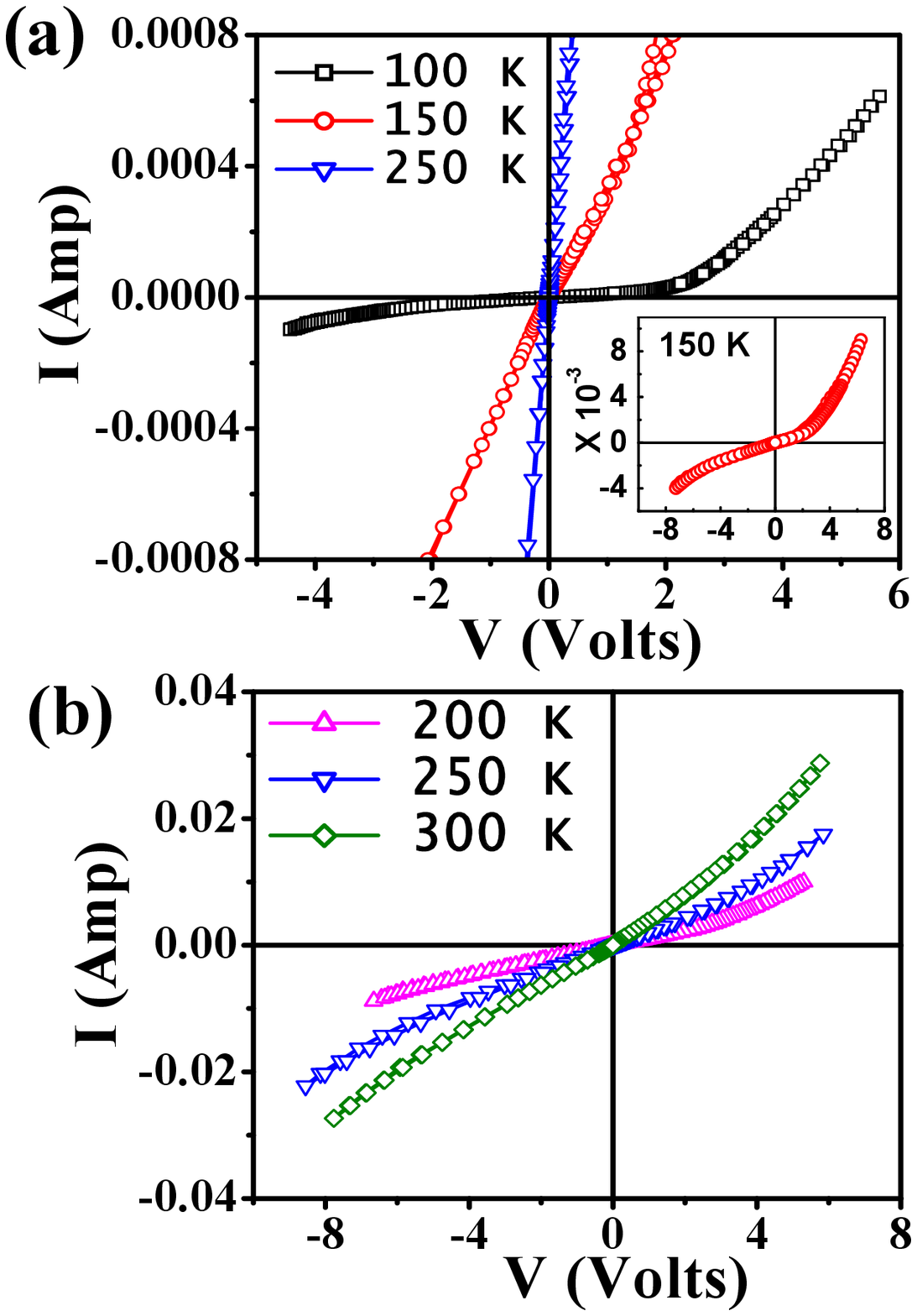}
	\caption{(Color online) (a) Current-Voltage (\textit{I-V}) characteristics of $Fe_{3}O_{4}$(111)/$Alq_{3}$/Co/Au SV structure at different temperatures. A drastic change in conduction mode with temperature can be clearly identified. Inset shows the \textit{I-V} curve at T = 150 K. (b) \textit{I-V} curves at T = 200, 250 and 300 K, exhibiting tunneling characteristics.}
	\label{fig:Fig1}
\end{figure*}

\begin{figure*}
	\centering
		\includegraphics[width=12cm]{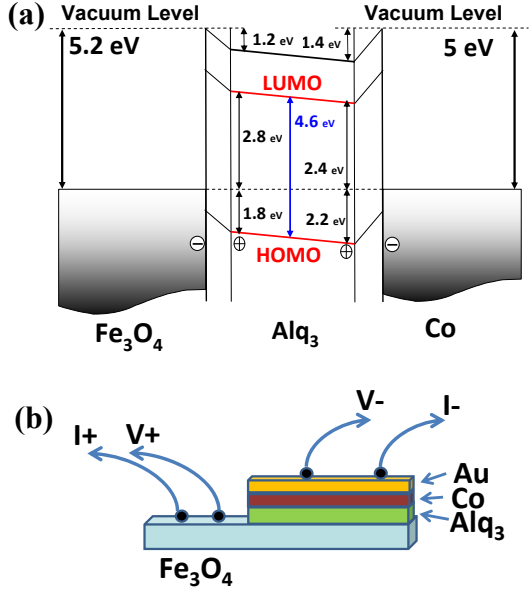}
	\caption{(Color online) (a) Energy level diagram at the interfaces of $Alq_{3}$ and $Fe_{3}O_{4}$/Co electrodes. The values of work functions of $Fe_{3}O_{4}$/Co electrodes and the transport gap of $Alq_{3}$ have been taken from literature. Shift of energy levels of $Alq_{3}$ with respect to vacuum level due to the formation of strong interface dipole is also indicated for both interfaces. For $Fe_{3}O_{4}$ and Co electrodes the values are 1.2 and 1.4 eV, respectively (from literature). (b) Schematic view of device and transport measurement geometry.}
	\label{fig:Fig2}
\end{figure*}

\begin{figure*}
	\centering
		\includegraphics[width=12cm]{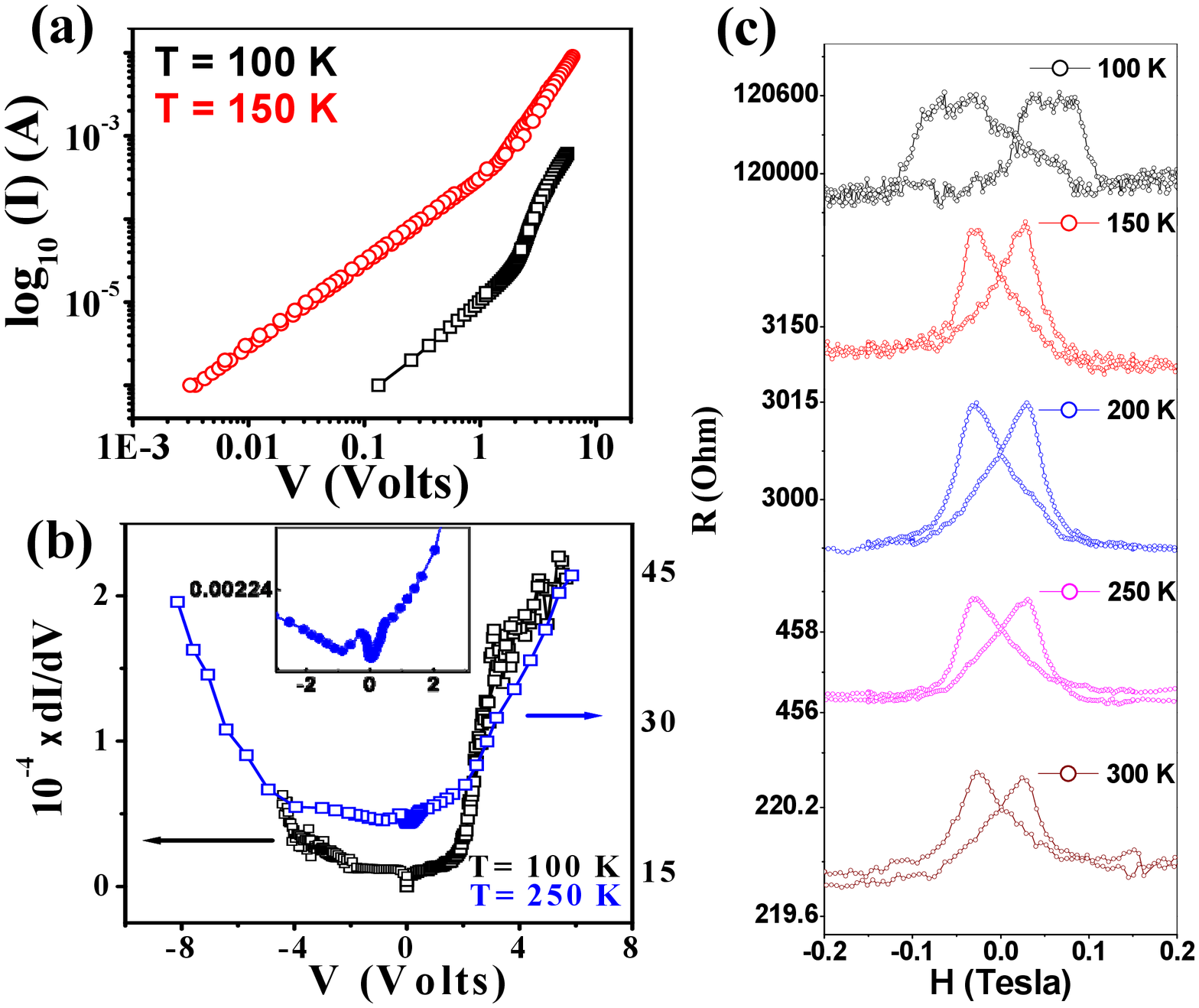}
	\caption{(Color online) (a)\textit{log I vs. log V} plots at T = 100 and 150 K. (b) Differential conductance vs. bias voltage at T = 100 and 250 K. Data are plotted on two different y scales, left and right y axis and the corresponding y axis is assigned by an arrow. Inset shows the zero bias anomaly at T = 250 K. (c) MR effect of the organic SVs. MR curves at different temperatures from 100 to 300 K. Noticeably, SV effect persists up to room temperature.}
	\label{fig:Fig3}
\end{figure*}

\begin{figure*}
	\centering
		\includegraphics[width=12cm]{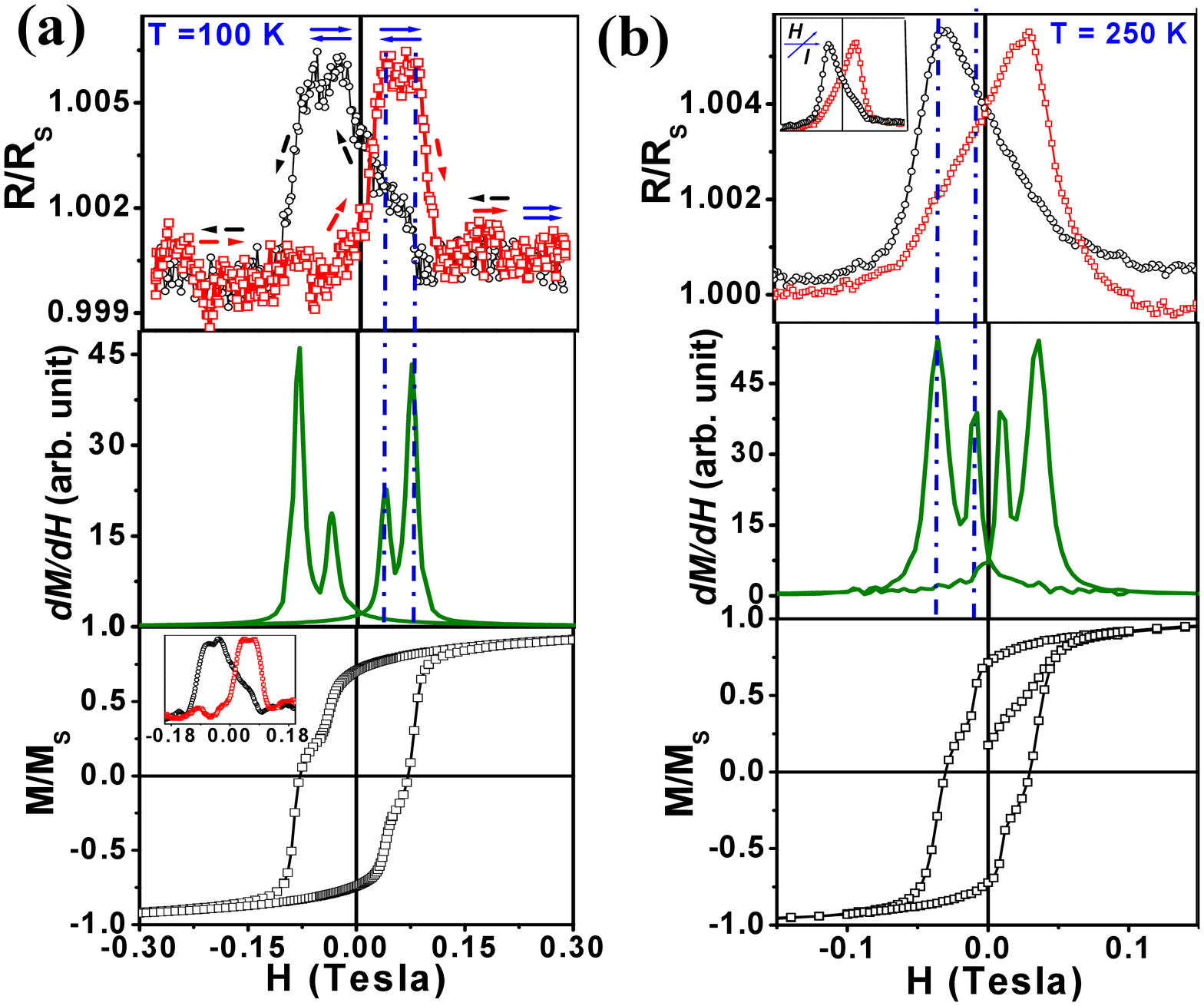}
	\caption{(Color online) MR, represented by resistance (R) normalized by the resiatance value ($R_{S}$) when two electrodes attain magentic satiuration [upper panel],derivative of magnetization with respect to magnetic field (dM/dH) [middle panel] and magnetization, normalized by saturation magnetization ($M_{S}$) [lower panel] hysteresis loops as a function of magnetic field (H), measured at (a) T = 100 K and (b) T = 250 K. Blue solid arrows [upper panel of (a)] are used to represent parallel and anti-parallel magnetization configurations. The dashed arrows in R/($R_{S}$) vs. H plot [upper panel of (a)] indicates the magnetic field sweep direction. Red and black colors are used for the arrows to express the correspondence of the MR curve with the magnetic field sweep direction. Inset in lower panel of (a) shows smoothened R/($R_{S}$) vs. H hysteresis loop at T = 100 K. Inset in upper panel of (b) shows the in-plane transverse R/($R_{S}$) vs. H hysteresis loop at T = 250 K.}
	\label{fig:Fig4}
\end{figure*}

\begin{figure*}
	\centering
		\includegraphics[width=12cm]{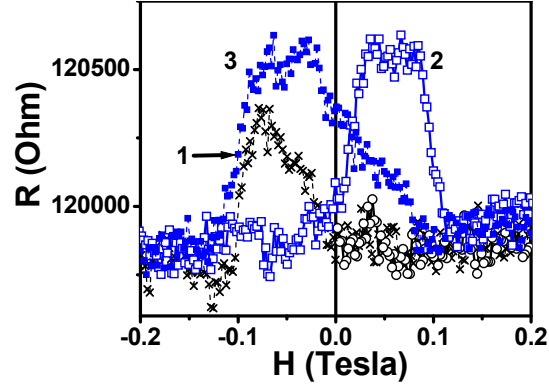}
	\caption{(Color online) MR hysteresis loop as a function of magnetic field, measured at T = 100 K for three cycles of magnetic field sweeping between positive to negative saturation magnetic field (where two electrodes attain magentic satiuration). Black open circle symbol corresponds initial magnetic field sweeping from H = 0 Tesla to saturation magnetic field in positive direction. Black cross symbol corresponds to the successive magnetic field sweeping from positve to negative saturation field. We denote this MR curve by 1. Blue open and filled square symbol corresponds to the consecutive magnetic field sweeping from negative to positive (denoted by 2) and positive to negative (denoted by 3) saturation magnetic field, respectively.}
	\label{fig:Fig5}
\end{figure*}

\begin{figure*}
	\centering
		\includegraphics[width=12cm]{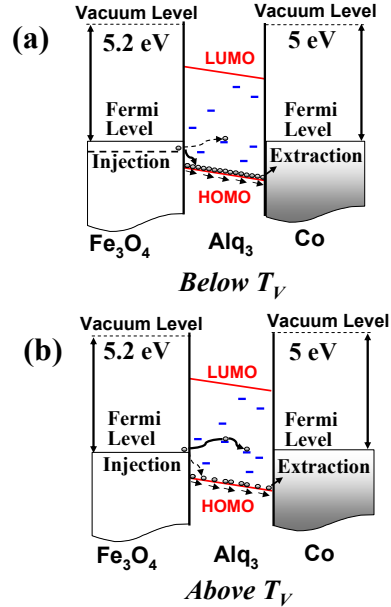}
	\caption{(Color online) Change in conduction mode from carrier injection to tunneling across Verwey transition temperature of $Fe_{3}O_{4}$ electrode. Schematic illustration of device operation at very low bias, (a) below and (b) above Verwey transition temperature.}
	\label{fig:Fig6}
\end{figure*}
\end{document}